\documentstyle[amsfonts,epsfig,12pt]{article}

\def\mypagenumber{1}

\def\myend{\end{document}}

\normalsize

\newcounter{sxn}

\newcounter{axn}

\date{}

\newdimen\mybaselineskip
\newcommand{\beeq}{\begin{equation}}
\newcommand{\eneq}{\end{equation}}
\newcommand{\be}{\begin{eqnarray}}
\newcommand{\ee}{\end{eqnarray}}
\newcommand{\bpic}{\begin{picture}}
\newcommand{\epic}{\end{picture}}

\def\dd{\partial}
\def\la{\raise.16ex\hbox{$\langle$} \, }
\def\ra{\, \raise.16ex\hbox{$\rangle$} }

\def\psibar{ \psi \kern-.65em\raise.6em\hbox{$-$} }
\def\mbar{ m \kern-.78em\raise.4em\hbox{$-$}\lower.4em\hbox{} }

\def\ep{\epsilon}

\def\n@space{\nulldelimiterspace=0pt \mathsurround=0pt }
\def\huge#1{{\hbox{$\left#1\vbox to 20.5pt{}\right.\n@space$}}}

\def\myskip{\noalign{\kern 8pt}}
\def\myeqspace{\noalign{\kern 10pt}}

\def\boxit#1{$\vcenter{\hrule\hbox{\vrule\kern3pt
    \vbox{\kern3pt\hbox{#1}\kern3pt}\kern3pt\vrule}\hrule}$}
\def\bigbox#1{$\vcenter{\hrule\hbox{\vrule\kern5pt
     \vbox{\kern5pt\hbox{#1}\kern5pt}\kern5pt\vrule}\hrule}$}

\def\ignore#1{{}}

\begin{document}
\bibliographystyle{unsrt}
\footskip 1.0cm

\thispagestyle{empty}
\setcounter{page}{\mypagenumber}

             
\begin{flushright}{
BRX-TH-499\\}

\end{flushright}

\vspace{2.5cm}
\begin{center}
{\LARGE \bf {Massive, Topologically Massive, Models}}\\ 
\vskip 1 cm
{\large{S. Deser and Bayram  Tekin  }}\footnote{e-mail:~
deser, tekin@brandeis.edu}\\
\vspace{.5cm}
{\it Department of  Physics, Brandeis University, Waltham, MA 02454,
USA}\\

\end{center}

\vspace*{1.5cm}


\begin{abstract}
\baselineskip=18pt
In three dimensions, there are two distinct mass-generating mechanisms 
for gauge fields: adding the usual Proca/Pauli-Fierz, 
or the more esoteric Chern-Simons (CS), terms. Here, we analyze the 
three-term models where both types are present, and their various limits.
Surprisingly, in the tensor case, these seemingly innocuous 
systems are physically unacceptable. If the sign of the Einstein term
is ``wrong'', as is in fact required in the 
CS theory, then the excitation masses are
always complex; with the usual sign, there is a (known) region of the two mass
parameters where reality is restored, but instead we show that 
a ghost problem arises,
while, for the ``pure mass '' 
two-term system without an Einstein action, 
complex masses are unavoidable.
This contrasts with the smooth behavior of the corresponding vector models.
Separately, we show that the 
``partial masslessness'' exhibited by (plain) massive
spin-2 models in de Sitter backgrounds is shared by the three-term 
system: it also enjoys a reduced local gauge
invariance when this mass parameter is tuned to the cosmological constant.
\end{abstract}

\vfill


\newpage

\normalsize
\baselineskip=22pt plus 1pt minus 1pt
\parindent=25pt
\vspace{1.5 cm}

Topologically massive tensor (TMG) gauge
theories in $D= 3$ are well-understood models, whose linearized versions
describe single massive but gauge-invariant excitations \cite{deser1}.
We analyze here the augmented, 3-term, system (MTMG) 
that breaks the invariance
through an explicit mass term; the vector analogs 
are briefly reviewed for contrast.

We are motivated by two quite separate developments: 
In the first, it was shown from the propagator's 
poles that the 
mass eigenvalues of a particular class of MTMG systems 
are solutions of a 
cubic equation, two of whose roots
become complex in a range of the underlying two ``mass'' parameter space
\cite{pinheiro}.
In contrast, the vector system's masses 
solve a quadratic equation with everywhere real, positive roots 
\cite{pisarski}. 
We will analyze the excitations
and mass counts for generic signs and values of these parameters and for 
both permitted sign choices of the Einstein action, as well as in its
absence. We will uncover not only the objectionable complex masses, 
but also exhibit the unavoidable presence of ghosts and tachyon 
excitations in the underlying two-term models. 

Our second topic is the study of MTMG in a constant curvature, rather than 
flat, background. Here we follow recent results which discuss the 
``partial masslessness'' of ordinary massive gravity at a value of the mass tuned to the cosmological constant, where a residual gauge invariance eliminates 
the helicity zero mode but also leads to non-unitary regions in 
the $(m^2, \Lambda)$ plane \cite{waldron}. It is  natural to ask whether this phenomenon persists for MTMG ( it has no vector analog), given the common gauge covariance of the two systems' kinetic terms, and we will see that it does.

The action we consider here is the sum of Einstein, (third
derivative order) Chern-Simons, and standard 
Pauli-Fierz mass, terms.
 \be
 I =\int_M d^3 x \, \left \{ a \sqrt{g} R -{1\over 2 \mu}\epsilon^{\lambda \mu \nu}\Gamma^\rho\,_{\lambda \sigma} 
( \partial_\mu \Gamma^\sigma\,_{\rho \nu} 
+ {2\over 3}\Gamma^\sigma\,_{\mu \beta}\Gamma^\beta\,_{\nu \rho} ) - 
{m^2 \over 4} (h_{\mu \nu}h^{\mu \nu} - h^2 ) \right \},  
\label{tmgaction}
 \ee
at quadratic order in $h_{\mu \nu} 
\equiv g_{\mu \nu} - \eta_{\mu \nu}$, $h \equiv \eta^{\mu \nu}h_{\mu \nu}$; 
our signature is $(-,+,+)$. Here the sign of $\mu$ is arbitrary 
but effectively irrelevant, that of $m^2$ is {\it{a priori}} free, 
while $a$ allows for 
choosing the Einstein term's sign ($a= +1$ is the usual one ) or even 
removing it, 
so this is the most general such model. 
All operations are with respect to the flat background $\eta_{\mu \nu}$. Note
that $\mu = \infty$ represents massive gravity with 2 excitations
(massive spin 2 in $3D$ has as many modes as massless spin 2 
in $4D$); $m = 0$ is TMG with 1 mode. 
Pure Einstein theory, $\mu^{-1} = m^2 =0$, has no excitations in D=3. 

At this point, one can already see one insurmountable
discontinuity latent in (\ref{tmgaction}): As was shown in \cite{deser1}, 
the sign of the Einstein term in pure TMG must be $a= -1$, 
opposite to that in the
usual Einstein gravity, in order for the energy to be positive,
independent of the sign of $\mu$. However since the usual massive spin-2 
system does have excitations, 
both the relative and overall signs of the Einstein and mass terms 
are forced to be the {\it
usual} Einstein and $m^2$ signs to avoid ghosts and
tachyons: there is an unavoidable conflict in the choice of
Einstein action sign $a$ in the two cases.

For comparison, we first describe the generic vector case with a Proca mass
term added to the TME action,
 \be
I= \int d^3x \left \{{- a \over 4} F_{\mu\nu} F^{\mu\nu}
+ {\kappa\over 2} \ep^{\mu\nu\rho} A_\mu \dd_\nu A_\rho -{1\over 2} m^2 
A_\mu A^\mu \right \}.
\label{tmeaction}
\ee
To begin with, one may set $a= +1$ as it must be positive both in 
the Proca ($\kappa = 0$ ) 
and TME ($m = 0$) limits; $a= -1$ would also introduce tachyons. 
[ As was noted long ago \cite{jackiw}, 
setting $a = 0$ yields just another version of TME and 
hence it is equivalent to setting $m^2 = 0$;
this also becomes clear in our canonical analysis of the 
theory.] Making a ``2+1" decomposition of the conjugate variables,  
\be
F_{0i}= \epsilon^{ij}\hat{\partial}_j \omega + \hat{\partial}_i f, \hskip 1cm
A_i= \epsilon^{ij}\hat{\partial}_j \chi + \hat{\partial}_i \varphi,
\hskip 1 cm \hat{\partial}_i \equiv {\partial_i \over \sqrt{-\nabla^2}},
\ee
reduces (\ref{tmeaction}) to 
\be
I= \int d^3 x\Big\{(\omega &+&\kappa\varphi) \dot{\chi} + f\dot{\varphi} -
{1\over 2}[ \chi ( m^2 -\nabla^2)\chi + \omega^2 + f^2 + m^2\varphi^2] 
\nonumber \\ 
  &+&{m^2\over 2} A_0^2 + A_0\sqrt{-\nabla^2 }(\kappa \chi- f) \Big\},
\ee
which, after eliminating $A_0$, can be diagonalized to represent 
two massive degrees of freedom with masses 
$m_\pm  = {1\over 2} \, \left\{ \sqrt{ \kappa^2
+ {4 m^2}} \pm |\kappa| \right\}$
in agreement with those found in the propagator poles \cite{pisarski}.
It is easily seen that the actions of various limiting theories are 
smoothly reached, including the equivalence of the $a = 0$ and    
the ``self-dual'' model of \cite{jackiw}.

To analyze MTMG  for generic values of ($\mu , m$), and
 the sign factor $a$ in (\ref{tmgaction}), we first   
decompose $h_{\mu \nu}$,   
\be
h_{ij} = (\delta_{ij} +{\hat{\partial}}_i {\hat{\partial}}_j )\,\phi - 
{\hat{\partial}}_i {\hat{\partial}}_j \chi + 
(\epsilon_{ik}\hat{\partial}_k\hat{\partial}_j
+ \epsilon_{jk}\hat{\partial}_k\hat{\partial}_i\,) \xi, \hskip 0.5 cm   
h_{0i}= -\epsilon_{ij}{\partial}_j \eta + \partial_i N_L, \hskip 0.5 cm
h_{00} \equiv N.
\ee
The various components of (\ref{tmgaction}) are 
\be
I_E = {a\over 2} \int d^3 x
\left \{\phi \ddot{\chi} + \phi \nabla^2 \,( N+ 2\dot{N_L} ) +
( \nabla^2 \eta +\dot{\xi})^2 \right \},
\ee

\be
I_{CS} = {1\over 2\mu}\int d^3 x \left \{
( \nabla^2 \eta +\dot{\xi} )\, [ \nabla^2 ( N+ 2\dot{N}_L ) + \ddot{\chi} +
\Box{\phi} ] \right \},
\ee

\be
I_{PF}=  {m^2\over 2} \int d^3 x \left \{ [ N_L\nabla^2 N_L + 
\eta \nabla^2 \eta 
+\xi^2 -\phi \chi+ N(\phi + \chi) ] \right \}.
\ee
Unlike the vector model, this generically represents three (rather than two) 
massive excitations: three of the six $h_{\mu \nu}$ can be eliminated 
by constraints. We can determine the mass spectrum without
having to diagonalize the fields, by forming the (cubic) eigenvalue equation,
\be
(\Box^3 - \mu^2 a^2\Box^2 + 2 a m^2 \mu^2 \Box - \mu^2 m^4 ) = 0.
\label{cubic}
\ee
This equation, for $a= 1$, was 
obtained from the pole of the propagator in \cite{pinheiro},
where it was also noted that the roots  are complex unless the mass 
parameter ratio $\mu^2/m^2 \equiv \lambda \geq 27/4$.  
The explicit form of the three roots is not particularly 
illuminating, but we note, for example, that at $\lambda = 27/4$ (for $a =1$) 
hitherto complex roots coalesce and the masses are simply
\be
m_1 = m_2 = 2m_3= {2\over 3}\mu = m \sqrt{3} 
\ee
The limit $a =0 $, corresponds to keeping only the CS and mass
terms, while dropping the Einstein part.  Here the eigenvalue
equation says that the 3 roots  are just $(|\mu| m^2)^{1/3}$ times
the cube roots of unity, whose two imaginary values are unavoidable:
this model is never viable. [ There is no tensor analog 
of  ``self-dual''-TME equivalence.]

The analysis so far has dealt with $a = +1$. 
However, the viability of the theory requires
two correct signs: the first one to avoid tachyons ,the second to avoid
ghosts, {\it{i.e.}} one needs both the relative sign in $(\Box - m^2)$ as well
as the overall sign in the action, +$ \int \phi (\Box -m^2) \phi$. 
In \cite{deser1}, it was shown that TMG required $a =-1$ for ghost-freedom 
( no tachyons arise for either sign choice.) Thus we conclude that
at least the small $m^2 $ limit to a two-term theory is unphysical, despite
having a real mass $\mu$ ( the seemingly massless other two modes are 
non-propagating). [We have not pursued in detail the diagonalization 
required to check the finite region for ghost signs.]
Indeed, for $a= -1$, there are two complex 
mass roots of (\ref{cubic}) for {\it{any}} finite $\mu^2/m^2$ value.
In summary, for our three-term models, $a= +1$ has acceptable mass ranges
but faces ghost problems, and both  $a=-1$ and $a = 0$ are 
always forbidden. None of 
these obstacles are present in the vector
models, thanks to their lower derivative order and quadratic mass roots.

Consider now the different issue of the behavior of our models in 
de Sitter ($\Lambda > 0$) backgrounds. In any dimension, 
massive gravity (or any other higher spin ) acquires gauge invariance 
at a non-zero mass parameter tuned to $\Lambda$ \cite{waldron}. 
The existence, in $ D= 3$, of the gauge invariant 
Cotton tensor, which gives mass to
spin-2 fields while keeping gauge invariance, warrants a separate 
discussion of MTMG in a cosmological background $g_{\mu \nu}$.
The latter is defined by
\be
R_{\mu \rho \nu \sigma} = \Lambda ( g_{\mu \nu} g_{\rho \sigma} -  
g_{\mu \rho} g_{\nu \sigma} ), \hskip 1 cm R_{\mu \nu} = 2\Lambda g_{\mu \nu},
\hskip 1cm R= 6\Lambda. 
\ee
The linearized Ricci tensor in this background reads
\be
R^L_{\mu \nu} = {1\over 2} \left\{ - \Box  h_{\mu\nu} - 
\nabla_\mu \nabla_\nu h  +   
\nabla^{\sigma} \nabla_\nu h_{\sigma \mu} +
\nabla^{\sigma} \nabla_\mu h_{\sigma \nu} \right \}.
\label{linearricci}
\ee
The unique, conserved,  ``Einstein'' tensor 
is
\be
G^L\,_{\mu \nu} \equiv  R^L\,_{\mu\nu} - {1\over 2} 
g_{\mu \nu}R_L - 2\Lambda ( h_{\mu\nu}- {1\over 2} g_{\mu \nu} h), 
\hskip  1 cm \nabla_{\mu}\,G_L\,^{\mu \nu} \equiv 0.
\ee
Let us recall \cite{waldron} that massive gravity,
\be
G^L\,_{\mu \nu} + m^2( h_{\mu \nu} - g_{\mu \nu}h ) = 0,
\label{spin2}
\ee
together with the Bianchi identities, leads (at $D= 3$) to  
the on-shell restriction
\be
m^2(\Lambda - 2 m^2 )h = 0.
\label{partmass}
\ee
For arbitrary $m^2$ and $\Lambda$, this just constrains the massive field
to be traceless. For $m^2 = \Lambda/2$,  a  ``partial masslessness'' arises
from the novel scalar gauge invariance at this point.

The unique linearized version of the Cotton tensor of TMG \cite{deser1} 
\be
C_L\,^{\mu \nu}= {1\over \sqrt{g}}\epsilon^{\mu\alpha \beta}\,\,
g_{\beta \sigma} \nabla_{\alpha}\,\left \{ 
{{R}}_L\,^{\sigma \nu} 
- 2\Lambda h^{\sigma \nu} - {1\over 4} g^{\sigma \nu} 
( R_L - 2\Lambda h ) \right \},
\ee
is still symmetric, traceless
and conserved with respect to the background,
\be
\epsilon^{\sigma \mu \nu}C^L_{\mu \nu}=0, \hskip 1cm 
g^{\mu \nu}C^L\,_{\mu \nu} =0 , \hskip 1 cm \nabla_\mu C^{L\,\mu \nu}=0.  
\ee
Hence MTMG in de Sitter space reads
\be
G^L\,_{\mu \nu} + {1\over \mu} C^L\,_{\mu \nu} + m^2( h_{\mu \nu} - g_{\mu \nu}h ) = 0.
\label{fullspin2}
\ee
Since it is traceless, the presence of $C_{\mu \nu}\,^L$ does not affect the 
the condition (\ref{partmass});
it merely adds one  degree of freedom to those of the massive theory.
The $(m^2, \Lambda)$ plane is divided into two regions by the 
same ($\mu$-independent ) $m^2 = \Lambda/2$ line for both \{ MTMG ,MG\}. 
In the $m^2 > \Lambda/2$ region there
are respectively $\{3, 2\}$ excitations. 
On the line, the helicity zero one vanishes leaving $\{2,1\}$ excitations. 
The $m^2 < \Lambda/2$ region is non-unitary due to the return of the helicity 
zero excitation with a non-unitary sign. Finally at $m^2=0$, we 
recover the full linearized diffeomorphism invariance 
and  therefore there are $\{1,0\}$ modes: this is just the 
\{TMG , free Einstein \} point.

We thank A. Waldron for discussions. This work was 
supported by the National Science Foundation Grant PHY99-73935.

\myend


\begin{thebibliography}{99}


\bibitem{deser1}
S.~Deser, R.~Jackiw and S.~Templeton,
Annals Phys.\  {\bf 140}, 372 (1982);
Phys.\ Rev.\ Lett.\  {\bf 48}, 975 (1982).

\bibitem{pinheiro}
C.~Pinheiro, G.~O.~Pires and N.~Tomimura,
Nuovo Cim.\ B {\bf 111}, 1023 (1996).

\bibitem{pisarski}
R.~D.~Pisarski and S.~Rao,
Phys.\ Rev.\ D {\bf 32}, 2081 (1985).
P.~N.~Tan, B.~Tekin and Y.~Hosotani,
Phys.\ Lett.\ B {\bf 388}, 611 (1996); See also 
R.~Banerjee, B.~Chakraborty and T.~Scaria,
Int.\ J.\ Mod.\ Phys.\ A {\bf 16}, 3967 (2001).

\bibitem{waldron}
S.~Deser and A.~Waldron,
Phys.\ Rev.\ Lett.\  {\bf 87}, 031601 (2001); 
Nucl.\ Phys.\ B {\bf 607}, 577 (2001).

\bibitem{jackiw}
S.~Deser and R.~Jackiw,
Phys.\ Lett.\ B {\bf 139}, 371 (1984).

\end{thebibliography}
\end{document}